\documentclass[prl,
 reprint,
 amsmath,amssymb,
 aps, 
 physrev,
]{revtex4-2}

\usepackage{graphicx}
\usepackage{bm}
\usepackage{soul}
\usepackage{graphicx}
\usepackage{textgreek}
\usepackage{xcolor} 
\usepackage{cancel}
\usepackage{siunitx}
\usepackage{layouts}
\usepackage[colorlinks]{hyperref}
\hypersetup{
    colorlinks=true,
    linkcolor=blue,
    filecolor=blue,      
    urlcolor=blue,
    citecolor=blue
}

\begin{document}

\preprint{APS/123-QED}

\title{\textbf{Loopless Multiterminal Quantum Circuits at Odd Parity} 
}%

\author{Antonio Manesco$^{1, 2}$}
\email{am@antoniomanesco.org}
\author{Anton Akhmerov$^2$}%
\author{Valla Fatemi$^3$}
 \email{vf82@cornell.edu}
\affiliation{%
    $^{1}$Center for Quantum Devices, Niels Bohr Institute, University of Copenhagen, DK-2100 Copenhagen, Denmark\\
    $^{2}$Kavli Institute of Nanoscience, Delft University of Technology, 2600 GA Delft, The Netherlands\\
    $^{3}$School of Applied and Engineering Physics, Cornell University, Ithaca, New York 14853, USA
}%


\date{May 28, 2026}

\begin{abstract}
We theoretically investigate loopless multiterminal hybrid superconducting devices at odd fermion parity with time-reversal symmetry.
We find that the energy-phase relationship has a double minimum corresponding to opposite windings of the superconducting phases. 
Spin-orbit coupling adds multi-axial spin splittings, which contrasts with two-terminal devices where spin dependence is uniaxial.
Capacitive shunting localizes quantum circuit states in the wells and exponentially suppresses their splitting.
For weak spin-orbit strength, the system has a four-dimensional spin-chirality low-energy subspace which can be universally controlled with electric fields only.

\end{abstract}

\maketitle


\section{\label{sec:intro}Introduction}

Andreev bound states hosted in Josephson devices offer the prospect of strong coupling between microscopic and macroscopic quantum degrees of freedom~\cite{janvier_coherent_2015}. 
The microscopic component is given by the spin or quasiparticle excitations hosted in the Andreev levels, and the macroscopic components are the microwave-frequency plasma modes of superconducting circuits. 
The coupling strength is derived from the magnitude of the supercurrent mediated by the Andreev states and the direct, galvanic coupling between the weak link and the circuit~\cite{devoret_circuit-qed_2007}.
In particular, Andreev spin states offer spin-dependent supercurrent due to spin-orbit coupling (SOC) in the weak link, which results in spin-circuit couplings that enable efficient readout and long-range spin-spin interactions~\cite{hays_continuous_2020,hays_coherent_2021,pita-vidal_strong_2024}.
In the future, long-range all-to-all interactions and disjoint support between single-spin and many-spin states may be useful for quantum information applications~\cite{pita-vidal_blueprint_2025,lu_kramers-protected_2025,kurilovich_andreev_2025}.

\begin{figure}[b]
    \centering
    \includegraphics[width=0.9\linewidth]{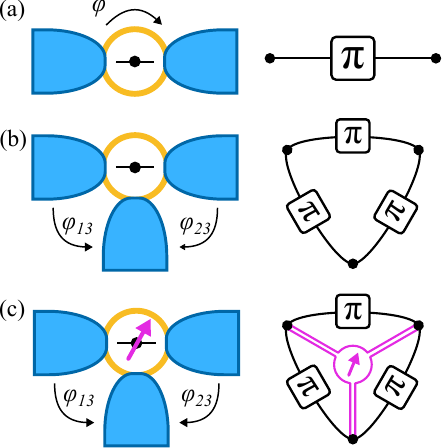}
    \caption{Schematics for proximitized single-level dots with two and three terminals in the odd parity sector and its analogous circuits.
    (a) Two-terminal devices without SOC are typically $\pi$-junctions.
    (b) Three-terminal devices without SOC are analogous to a triangle of $\pi$-junctions at zero flux, which hosts a double-well, spin-independent energy-phase relation.
    (c) Three-terminal devices with SOC include spin-dependent contributions spanning the full SU(2) structure of the spin to the spin-phase energy relation, alongside the spinless triangle of $\pi$-junctions.}
    \label{fig:intro}
\end{figure}

The basic physics of two-terminal Andreev spins is captured by the Josephson coupling~\cite{chtchelkatchev_andreev_2003,padurariu_theoretical_2010,pavesic_generalized_2024}
\begin{equation}
    U_\mathrm{2T} = E_0 \cos\varphi + E_\sigma \sigma_z \sin \varphi ~,\label{eq:two_terminal}
\end{equation}
where $\varphi$ is the phase drop across the weak link, and $\sigma_z$ is a Pauli matrix for the spin degree of freedom.
$E_0>0$ is typical for odd-parity, single-channel weak links, \emph{i.e.}, they behave as $\pi$-junctions in the spinless case (Fig.~\ref{fig:intro}(a))~\cite{bulaevskii_superconducting_1977,van_dam_supercurrent_2006,fatemi_microwave_2022}.
The spin-dependent part of the spin-phase energy relation has approximately a single Pauli operator in the low-energy description.
This is tied to the electrons in the weak link following effectively a single path between the superconducting reservoirs.

Multiterminal devices have been proposed as a route to achieving large spin splittings and controlling fermion parity transitions~\cite{van_heck_single_2014,svetogorov_theory_2025}. 
Both effects were recently observed in tunneling experiments~\cite{coraiola_spin-degeneracy_2024}.
This suggests that integration of multiterminal Andreev devices in dynamical microwave circuits is viable, but multiterminal Andreev states in dynamical circuits have so far been explored theoretically in the even parity sector~\cite{matute-canadas_quantum_2024} or without spin-orbit coupling and with loops~\cite{feinberg_spontaneous_2014}.

\begin{figure*}[t] 
    \centering
    \includegraphics[width=0.9\linewidth]{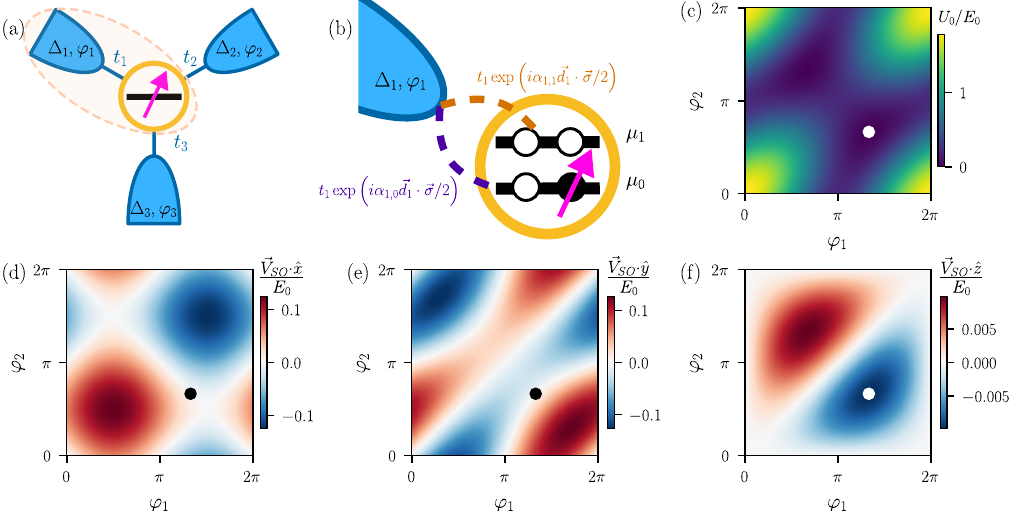}
    \caption{ 
    (a) Schematic diagram of the microscopic toy model for obtaining the Hamiltonian of the trijunction at odd parity. 
    (b) Schematic diagram of the structure of the dot levels and the tunneling terms that are included in the model. 
    (c) Spin-phase energy relation, $U = U_0 + U_{SO}$, for $t=0.1\Delta$, $\alpha=\pi / 30$, and $\mu_1=5\Delta$. 
    We set $\varphi_3 =0$ and so only refer to the phases of leads 1 and 2 on the axis. 
    (d-f) Spin-dependent part of the spin-phase energy relation, $U_{SO}$, for the terms proportional to (d) $\sigma_x$, (e) $\sigma_y$, and (f) $\sigma_z$.
    The black and white dots in panels (c-f) mark the location of the minimum of $U_0$.
    }
    \label{fig:EPRs}
\end{figure*}    

Here, we develop a minimal model of a quantum dot in the odd parity sector with tunnelings to more than two superconducting reservoirs, both with and without spin-orbit coupling. 
Without spin-orbit coupling, we find that the circuit is equivalent to a triangle of $\pi$-junctions~\cite{feinberg_spontaneous_2014} (Fig.~\ref{fig:intro}(b)), which exhibits a balanced double-well potential. 
Adding spin-orbit interaction then introduces a spin-phase energy relation (SPER) that involves all three Pauli vectors (Fig.~\ref{fig:intro}(c)). 
Finally, we introduce capacitive shunts and describe the spectra and inter-level matrix elements of the combined microscopic/macroscopic hybrid system. 
We highlight in particular that the loopless nature of this circuit does not admit a magnetic flux parameter, whereas double-well potentials in conventional Josephson junction circuits normally require loops and nonzero flux~\cite{orlando_superconducting_1999,doucot_pairing_2002,manucharyan_fluxonium_2009,gyenis_experimental_2021,schrade_protected_2022,hassani_inductively_2023}, which introduces noise sources that our circuit does not experience. 
Therefore, this approach has the potential to produce both heavy superconducting circuit qubits and Andreev spin qubits with noise resilience by avoiding a typical tradeoff.

\section{Spin-Phase Energy Relation of Multiterminal Dots at Odd Parity}

We present here a minimal model for a multiterminal weak link, treating the phases of the leads as classical variables for now.
This model is derived in the Supplementary Material, starting from a microscopic description and neglecting contributions from the continuum. 
Similar to previous works on two-terminal devices~\cite{padurariu_theoretical_2010, bargerbos_spectroscopy_2023}, we consider a two-orbital quantum dot with orbitals indexed by $j \in \{0, 1\}$, with corresponding chemical potential $\mu_j$, as depicted in Fig.~\ref{fig:EPRs}(a, b).
The presence of a second orbital is essential to account for cotunneling processes that lead to spin-orbit splitting of the resulting Andreev states.
The quantum dot is coupled to multiple superconducting leads with tunneling amplitudes $t_i$ from the dot to the lead $i$ with pairing amplitude $\Delta_i e^{i \varphi_i}$.
Moreover, every tunneling includes orbital-dependent spin-orbit effects with rotation angle $\alpha_{ij}$ about the Pauli vector $\vec{d}_i$.
Our model differs from the one developed by Svetogorov, \textit{et al.},~\cite{svetogorov_theory_2025} because it avoids the explicit use of a real-space loop, adding higher-orbital states instead.
This ensures that the spin degeneracy occurs only at the time-reversal symmetric points in the phase space.
We obtain the SPER up to the leading non-vanishing order in perturbation theory, consequently limiting the analytic expressions to the lowest harmonics of the phase difference $\varphi_{ij}=\varphi_i-\varphi_j$ across the superconducting leads $i$ and $j$.
We further separate the SPER $U = U_0 + U_{SO}$, into the spinless term $U_0$  and the spin-dependent term $U_{SO}$ caused by spin-orbit coupling.

The spinless part of the SPER in the weak-coupling limit ($t_i \ll \Delta_i, \mu_1$), $\emph{i.e.}$ when higher-harmonics are negligible, is the sum of $\pi$-junction potentials for each pair of leads,
\begin{equation}
    U_0  = \sum_{i,j} E_0^{ij} \cos \varphi_{ij}~, 
    \label{eq:spinless-potential}
\end{equation}
with $E_0^{ij} \sim \left[t_i^2 t_j^2 / (\Delta_i\Delta_j\mu_1)\right]\cos \theta_{ij}$, where $\theta_{ij}$ is the spin-dependent dynamical phase accumulated over the cotunneling process involving the higher energy dot level.
All the parameters are schematically shown in Fig.~\ref{fig:EPRs} for a three-terminal setup.
The potential \eqref{eq:spinless-potential} is the same as that of a triangle of $\pi$-junctions at zero flux~\cite{feinberg_spontaneous_2014}, schematically represented in Fig.~\ref{fig:intro}(b).
Thus, as long as all the $E_0^{ij}$ are comparable, and the spin-dependent terms are negligible, the energy--phase relation has the form of the symmetric double-well potential shown in Fig.~\ref{fig:EPRs}(c).
Note that we set $\varphi_3=0$ throughout so that $\varphi_{i3}=\varphi_i$.
The minima are located at points with opposite phase chirality due to time-reversal symmetry.

The spin-dependent term of the SPER is
\begin{equation}
    U_{SO} = \vec{V}_{SO}\cdot\vec{\sigma}~,\quad \vec{V}_{SO} = \sum_{i, j} E_\sigma^{ij}~\vec{n}_{ij} \sin \varphi_{ij}~,
    \label{eq:spinful-potential}
\end{equation}
where $E_\sigma = \left[t_i^2 t_j^2 / (\Delta_i\Delta_j\mu_1)\right] \sin \theta_{ij}$, and $\vec{n}_{ij}$ are unit vectors.
This differs from the two-terminal case in Eq.~\eqref{eq:two_terminal} because it requires more than one spin direction. 
The exact form of $\vec{n}_{ij}$ depends on microscopic details of the device.
Still, notably, it can include spin directions that are absent from the low-energy band structure of the normal material (see Supplemental Material for details).

We now consider a three-terminal setup with Rashba spin-orbit coupling in a two-dimensional electron gas.
This choice is similar to~\cite{svetogorov_theory_2025} and compatible with previous works on two-terminal quantum dot setups~\cite{bargerbos_spectroscopy_2023} and Rashba wires~\cite{park_andreev_2017} as well as existing multiterminal setups~\cite{coraiola_spin-degeneracy_2024}.
In two terminal devices, $E_0^{ij}$ and $E_{\sigma}^{ij}$ lies in the range of $0.1-\qty{1}{\giga\hertz}$~\cite{hays_coherent_2021, bargerbos_spectroscopy_2023, lu_andreev_2025}, and an order of magnitude larger in multiterminal devices~\cite{coraiola_spin-degeneracy_2024}.
The resulting SPER for a symmetric junction ($t_i = t$, $\Delta_i = \Delta$, $\alpha_{i,0} = \alpha$, and $\alpha_{i,1} = -\alpha$, $\vec{d}_i = (\cos \beta_i, \sin \beta_i)$, $\beta_i = 2\pi i / 3$, $E_\sigma^{ij}=E_{\sigma}$, and $E_0^{ij}=E_0$) is shown in Fig.~\ref{fig:EPRs}(d-f).
We observe that, if $E_\sigma \ll E_0$, the potential minima coincide with the point where the resulting spin-orbit field points along the $z$-direction.
The quantization axis, set by $\vec{V}_{SO}$, tilts away from $z$ with contributions from two different Pauli vectors for different trajectories in the phase space: $\sim(\varphi_1 + \varphi_2)\sigma_x,~(\varphi_1 - \varphi_2)\sigma_y$.
In an asymmetric device, the quantization axis is tilted away from the $z$-direction, and trajectories along $\varphi_+ = \varphi_1 + \varphi_2$ and $\varphi_- =\varphi_1 - \varphi_2$ tilt along linearly independent, rather than orthogonal, spin directions.
The phenomenology of the quantum circuits discussed below is unaffected by these quantitative changes.

\section{Quantum Circuits with Three-Terminal Dots at Odd Parity}

\begin{figure}
    \centering
    \includegraphics[width=1.0\linewidth]{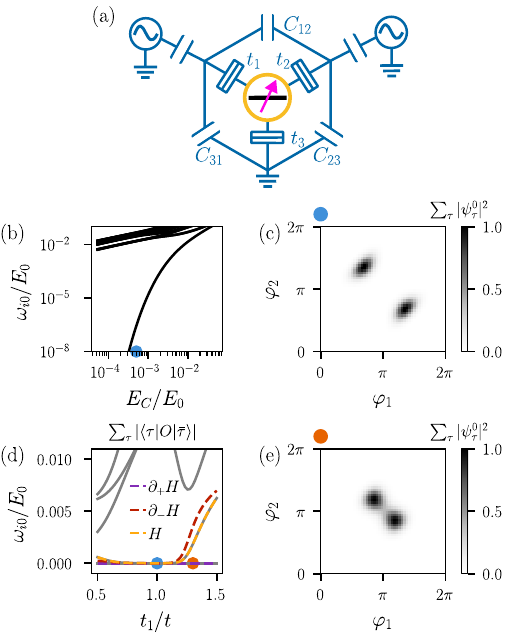}
    \caption{ 
    (a) Circuit diagram for a capacitively-shunted Andreev trijunction.
    (b) The circuit spectrum of the spinless case as a function of the circuit charging energy $E_C$ for $t=0.75\Delta$. The energy splitting between the lowest pair of states is exponentially suppressed as $E_C/E_0 \rightarrow 0$.
    (c) Probability density of the circuit wave functions in phase space for $E_C/E_0=5\times 10^{-4}$--indicated by the blue dot in panels (b) and (d).
    (d) The spectrum (gray lines) and matrix elements of the Hamilonian ($O=H)$ and linear corrections due to charge driving $O=\partial_{\pm}H$ for inter-chirality coupling (dashed lines) as a function of tuning one tunneling rate $t_1/t$, for $E_C/E_0 = 5\times 10^{-4}$--indicated by the blue dot in panel (b). 
    (e) Probability density of the circuit wave functions in phase space for $t_1 / t = 1.3$ and $E_C/E_0=5\times 10^{-4}$--indicated by the orange dot in panel (d).
    }
    \label{fig:spinless_circuit}
\end{figure}

Next, we shunt the circuit with capacitors, as shown in Fig.~\ref{fig:spinless_circuit}(a), so that the phase degrees of freedom have quantum dynamics; concomitantly we promote the phase operators to quantum operators $\left(\varphi_1, \varphi_2 \right)\rightarrow\left(\hat{\varphi}_1, \hat{\varphi}_2 \right)$. 
The capacitive parts of the circuit lead to a term of the form
\begin{equation}
    T = \frac{1}{2} E_C \vec{\hat{N}}^T \mathcal{C}^{-1}  \vec{\hat{N}}~,
\end{equation}
where $\vec{\hat{N}}^T = \left(\hat{N}_1, \hat{N}_2 \right)$ are the charge number operators conjugate to the phase operators $\left(\hat{\varphi}_1, \hat{\varphi}_2 \right)$, $E_C$ is a scale for the charging energy of the circuit (not of the dot), and $\mathcal{C}^{-1}$ is the normalized inverse of the capacitance matrix~\cite{vool_introduction_2017}. 
Expressions including offset charge are given in the appendix.
Combining this with the SPER gives the full circuit Hamiltonian $H=T+U$, which we diagonalized numerically, as described in the Supplementary Material and available in~\cite{rigotti_manesco_2026_18244491}.

\subsection{Spinless case}

For the spinless case ($\alpha=0$), we obtain the energy-phase relation from the full microscopic model shown in the Supplementary Material rather than restricting our analysis to the minimal model captured by Eqs.~\eqref{eq:spinless-potential} and~\eqref{eq:spinful-potential}.
With this choice, we can also take our analysis beyond the weak coupling limit and target the maximal depth of the double-well potential at $t\sim \Delta$.

First, we inspect the system while varying the circuit charging energy scale $E_C$, shown in Fig.~\ref{fig:spinless_circuit}(b).
As $E_C/E_0 \rightarrow 0$, the energy splitting between the lowest two levels drops exponentially (Fig.~\ref{fig:spinless_circuit}(b)).
Correspondingly, the wavefunctions are localized in the two wells (Fig.~\ref{fig:spinless_circuit}(c)). 
The phase expectation values for the ground states localized in each well correspond to current flows of different chirality.
We will describe this chirality as an effective quantum number $\tau$ characterizing the low-energy quantum states of the circuit. 

Next, we consider the behavior of the circuit while tuning one of the electron tunneling rates $t_1$ relative to the others $t$, \emph{e.g.}, via the electric field effect by adjusting the voltage on a gate electrode.
This is a typical way to activate exchange interactions between spin qubits~\cite{burkard_semiconductor_2023}.
Because time-reversal symmetry is preserved, the wells are always balanced, but they can approach each other or develop a lower barrier between them.
A similar strategy, but involving fast flux modulation, has been proposed as a method for control of flux qubits~\cite{PhysRevResearch.6.023064}.
Thereby, the energy splitting of the lowest two states can increase.
This evolution is shown in Fig.~\ref{fig:spinless_circuit}(d), with a corresponding probability density map of the lowest two states for a particular $t_1$ in Fig.~\ref{fig:spinless_circuit}(e). 

Finally, we consider the effect of an alternating voltage applied to a circuit node with a mediating capacitor. 
This is a typical way to excite plasma modes of superconducting qubit circuits~\cite{blais_circuit_2021}. 
We focus on the transition between the two lowest states, for which we find that the transition dipole matrix element rises exponentially from zero, as $t_1$ is varied, similar to the energy splitting. 
Figure~\ref{fig:spinless_circuit}(d) shows the expectation value of the matrix elements of $\partial_{\pm} H$ for transitions between the two states in the wells, with indices denoting derivatives with respect to $N_{\pm}=N_1 \pm N_2$, and $N_i$ is the offset charge in the island $i$.
Therefore, full control over the two-level subspace is possible by varying $t_1$ in combination with charge driving~\cite{zhang_universal_2021}. 
We remark that our approach can detect spontaneous emergence of chirality without introducing flux sensitivity or potentially picking up external sources of TRS breaking~\cite{feinberg_spontaneous_2014}.

\subsection{Spinful case}

\begin{figure}
    \centering
    \includegraphics[width=1.0\linewidth]{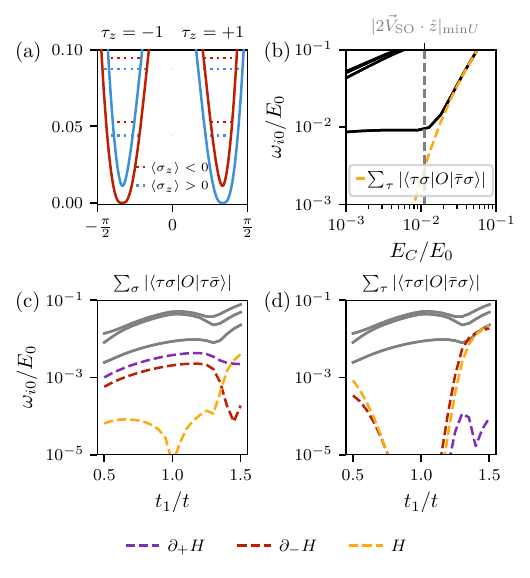}
    \caption{ 
    (a) The lowest eight energy levels for the heavy spinful circuit with weak SOC located in their respective energy wells, for $t=0.1\Delta$, $\alpha=\pi/20$, $\mu_1=5\Delta$, $E_C/E_0 = 10^{-3}$.
    (b) Spectrum as a function of $E_C/E_0$.
    Unlike the spinless case, the lowest available transition frequency (from the ground state doublet to the first excited doublet) saturates for $E_C < 2\vec{V}_{SO}\cdot \hat{z}\vert_{\min U}$.
    (c) Chirality-preserving, spin-flipping matrix elements as a function of $t_1/t$ plotted alongside the spectrum. Both drive orientations have a strong matrix element when the states are localized within the wells. These two drive orientations correspond to different Pauli matrices.
    The strength of the process is the same for both chiralities due to time-reversal symmetry.
    (d) Spin-preserving, chirality-flipping matrix elements, which find that the adiabatic coupling (yellow) and one off-diagonal matrix element (red) are nonzero only with appreciable detunings of $t_1/t$.
    The strength of the process is the same for both spins due to time-reversal symmetry.
    We used $E_C/E_0 = 10^{-3}$ in panels (c) and (d).
    }
    \label{fig:spinful_circuit}
\end{figure}

Incorporating the spinful part of the Hamiltonian, we find that the levels localized in the wells split into spin-orbit doublets, as shown in Fig.~\ref{fig:spinful_circuit}(a).
Here, for simplicity, we use the analytical expressions from Eqs.~\eqref{eq:spinless-potential} and~\eqref{eq:spinful-potential}.
Expressing the `chirality' degree of freedom with Pauli matrices $\tau$, we can write down a minimal effective Hamiltonian for this subspace as
\begin{equation}
    H_\mathrm{eff} = \frac{\epsilon}{2} \sigma_z \tau_z + \sum_j \delta N_j \left(\vec{g}_{\sigma, j}\cdot \vec{\sigma}~\tau_z + \sigma_0 \vec{g}_{\tau, j}
    \cdot \vec{\tau}\right)~,
    \label{eq:effective-qubit-hamiltonian}
\end{equation}
with the static term $\propto \epsilon$.
The other two terms of this Hamiltonian are activated via charge driving ($\delta N_j$) and generate spin-flipping, $(\vec{g}_{\sigma, j})_i = \operatorname{Tr}(\sigma_i \tau_z\partial_jH)$, and chirality-flipping, $(\vec{g}_{\tau, j})_i = \operatorname{Tr}(\sigma_0 \tau_i\partial_jH)$, processes.
We focus on the regime for which $\epsilon \sim E_\sigma$, valid when $E_\sigma \ll \omega_p \ll E_0$, where $\omega_p$ is the plasmon frequency.

The spin/chirality structure of the couplings in Eq.~\eqref{eq:effective-qubit-hamiltonian} is a direct consequence of the spin-dependent part of the SPER, shown in Fig.~\ref{fig:EPRs}(c-f): near the minimum of the potential wells, different trajectories in phase space correspond to different transverse spin rotations. 
Since these behaviors are time-reversed, opposite rotations occur for the same trajectories for opposite chirality.
For the symmetric configuration in particular, we find that driving with symmetric charge $\hat{N}_+ = \hat{N}_1+\hat{N}_2$ corresponds to an operator of the form $\sigma_x \tau_z$ and driving with antisymmetric charge $\hat{N}_- = \hat{N}_1-\hat{N}_2$ corresponds to an operator of the form $\sigma_y \tau_z$.

These chirality-dependent spin-flipping terms present an opportunity to explore the influence of circularly polarized driving of the two active nodes of the circuit.
We consider simultaneously driving the system with two voltages, $\Omega\sigma_x \tau_z \cos(\omega t)$ and $\pm \Omega\sigma_y \tau_z \sin(\omega t)$, to obtain 
\begin{equation}
    H_d = \Omega\left[\sigma_+ \exp(\pm i \omega t) + \mathrm{h.c.}\right]\tau_z~.
\end{equation}
Applying a unitary transformation $U = \exp\left( i H_\mathrm{eff} t/\hbar \right)$ to obtain $H_d' = U H_d U^\dagger$, and then applying a rotating wave approximation by assuming $|\hbar\omega - \epsilon|\ll \hbar\omega+\epsilon$, produces 
\begin{equation}
    H_d' = -\frac{\Omega}{2} \sigma_x (\tau_z \pm 1)~.\label{eq:chirality-selective-spin-rotation}
\end{equation}
This result indicates chirality-selective spin rotations are possible under resonant, circularly polarized driving. 

Field-effect tuning of the electron tunneling rates can also be used for chirality mixing. 
This may accomplish spin-independent inter-chirality mixing, proportional to $\tau_x$, as in the spinless case. 
Here, such (adiabatic) tuning may also induce phase accumulation for states with the same chirality, proportional to $\sigma_z\tau_z$. 
Both processes are shown together in Fig.~\ref{fig:spinful_circuit}(d). 
Combined with the intra-chirality spin rotations, this gives sufficient generators of SU(4) to accomplish universal control over this four-level subspace. 

\section{Discussion}

We first estimate the potential for such systems to be realized in the lab. 
For $t\sim\Delta$, the barrier between the wells is of the scale of $\Delta/6$. 
The ground state energy splitting therefore drops to the scale of \SI{1}{\hertz} for $E_C / h \approx \SI{1}{\mega\hertz}$ and $\Delta/h=\SI{45}{\giga\hertz}$ (similar to that of thin film aluminum).
For $\Delta$ comparable to the superconducting gap of niobium rather than aluminum, $E_C/h \approx \SI{10}{\mega\hertz}$ would be sufficient.
The spin-dependent contribution to the SPER ranges from $E_\sigma/h \sim 0.1-\SI{1}{\giga\hertz}$ in two-terminal devices~\cite{tosi_spin-orbit_2019,hays_continuous_2020,hays_coherent_2021,pita-vidal_direct_2023,pita-vidal_strong_2024,lu_andreev_2025} and was found to be as large as \SI{9}{\giga\hertz} in a three-terminal device~\cite{coraiola_spin-degeneracy_2024}. 
Therefore, we expect that the physics described here should be accessible.
We leave circuits with $E_\sigma \gtrsim E_0$ to be investigated in a future work. 

Both the spinless and spinful circuits have potential for use as qubits. 
In particular, they host a double-well potential without any loops, \emph{i.e.}, without any magnetic flux parameter.
In conventional Josephson circuits with tunneling junctions, magnetic flux is required to frustrate Josephson energies to produce a double-well potential~\cite{orlando_superconducting_1999,doucot_pairing_2002,manucharyan_fluxonium_2009,gyenis_experimental_2021,schrade_protected_2022,hassani_inductively_2023}.
This results in a tradeoff: heavier circuits, which suppress energy relaxation, become the most sensitive to flux noise~\cite{gyenis_moving_2021,hassani_inductively_2023}.
Our concept makes possible a new way forward for extremely heavy circuits to suppress energy relaxation without compromises to flux noise. 
This approach can use superconductor-semiconductor materials technology that admits control relying purely on the electric field effect rather than unconventional superconductivity~\cite{brosco_superconducting_2024}, at the cost of a resulting Josephson coupling bounded by the superconducting gap.
Similarly, the spinful circuit shows a way forward for Andreev spin qubits without using flux bias to split the spin degeneracy, for which flux noise would also ultimately limit coherence. 

We now identify several directions for future theoretical and experimental research.
First, we address extensions of the minimal models. 
The physics described here should be inspected again after adding physics to stabilize the odd parity sector, such as charging energy~\cite{pavesic_generalized_2024}. 
This approach has been used extensively with two-terminal Andreev spins~\cite{bargerbos_singlet-doublet_2022,bargerbos_spectroscopy_2023,pita-vidal_direct_2023}.
However, the physics here is sensitive to the tunneling rate, and it will be important to consider the Yu-Shiba-Rusinov regime~\cite{baran_surrogate_2023, zalom_hidden_2024}. 
A three-terminal interacting dot with superconducting leads has been briefly studied in a previous work~\cite{baran_surrogate_2023}, but its full phase diagram and effects of spin-orbit have not yet been calculated, to the best of our knowledge.
Furthermore, the quantum dynamics of the circuit may also renormalize, particularly in intermediate and high impedance environments~\cite{houzet_josephson_2024,gungordu_quantum_2025}.

Another direction is to more accurately account for the nonzero size of the dot and other aspects of real devices (\emph{e.g.}, electrostatic effects and actual material band structures). 
In our current approximation, the contribution producing the double-well potential grows slowly with the tunneling rate, but the physics of weak links can be nonanalytic in the size~\cite{levchenko_singular_2006,fatemi_nonlinearity_2025}.
Therefore, properly accounting for the effects of nonzero size and shape will help to clarify the depth of the double-well potential and the strength of the harmonics in the SPER for realistic devices.
Similarly, accounting for the actual band structure of realistic semiconductor hosts for the normal region (\emph{e.g.}, InAs~\cite{hays_coherent_2021}, InSb~\cite{zhang_evidence_2025}, and Ge~\cite{lakic_quantum_2025}), and their interaction with electrostatics and confinement, will be important for understanding the spin-dependent parts of the SPERs.

Finally, we expect that tiling such multiterminal devices in one- and two-dimensional arrays presents new directions.
Similarly to charge driving control, we expect that capacitive coupling between qubits would result in nontrivial two-qubit couplings.
A one-dimensional tiling of the spinless realization, for example, could produce a pseudospin chain that also protects a disjoint, doubly-degenerate ground state.
The spinful case with weak SOC has a higher-dimensional low-energy Hilbert space due to the four-level low-energy subspace, though strong SOC may return the system to a more strongly-coupled spin-chirality two-level subspace per trijunction. 
Then, in one- and two-dimensional arrays, we can envision long-range, non-collinear spin-spin interactions.
Appropriately tuned, these interactions could be used for quantum simulation of quantum phase transitions or spin liquid physics~\cite{dorier_quantum_2005,kitaev_anyons_2006}.
Finally, spin-spin interactions in a small multiterminal junction array could be useful for quantum error correction or quantum error protection against local spin noise.

\begin{acknowledgments}
We thank Chao-Ming Jian and Bernard van Heck for insightful discussions, Isidora Araya Day for remarks on the manuscript. V.F. thanks Colin Riggert for discussions and references regarding multiterminal YSR states. 

V.F. acknowledges that the research was sponsored by the Army Research Office and was accomplished under Grant Number W911NF-22-1-0053.
A.~M.~acknowledges the funding from the European Research Council (Grant Agreement \textnumero 856526).

\paragraph{Authors' contributions}
V.~F.~initiated and supervised the project.
V.~F.~and A.~A.~identified the equivalence between an odd-parity trijunction and three $\pi$-junctions.
A.~M.~developed the microscopic model.
The authors jointly developed the coherent control of the quantum states.
A.~M.~and V.~F.~performed the simulations and wrote the manuscript with input from A.~A.

\paragraph{Code and data availability}
The code and data used to prepare this manuscript are available in Zenodo~\cite{rigotti_manesco_2026_18244491}.

\paragraph{AI usage disclosure}
We have used OpenAI Codex with the GPT-5.2 Codex model to set up the Pymablock computation and proofread the text.

\end{acknowledgments}

\bibliography{references-vf, zenodo, more_refs}

\onecolumngrid
\newpage

\appendix

\section{\label{app:minimal-model} Minimal Model Derivation}

We consider a set of superconducting leads weakly coupled to a two-orbital normal dot/island with total Hamiltonian:
\begin{equation}
    H = \sum_n \Psi_{n}^{\dagger} H_{\text{island}}^n \Psi_{n} + \sum_i \Phi_i^{\dagger} H_{SC}^i \Phi_i + \sum_{i, n} \left(\Phi_i^{\dagger} H_T^{i,n} \Psi_n + h.c.\right)~,
    \label{eq:microscopic-model}
\end{equation}
where $H_{\text{island}}^n$ is the island Hamiltonian of the $n$-th orbital, $H_{SC}^i$ is the Hamiltonian of the $i$-th lead, and $H_T^{i,n}$ is the hopping from the $i$-th lead to the $n$-th orbital.
The Nambu spinors are $\Psi_n = (c_{n\uparrow}, c_{n\downarrow}, - c_{n\downarrow}^{\dagger}, c_{n\uparrow}^{\dagger})$, with $c_{n\sigma}$ the annihilation operator of an electron at orbital $n$ and spin $\sigma$, and $\Phi_i = (f_{i\uparrow}, f_{i\downarrow}, - f_{i\downarrow}^{\dagger}, f_{i\uparrow}^{\dagger})$, where $f_{i\sigma}$ are annihilation operators of electrons in the lead $i$ with spin $\sigma$.
Each orbital of the island contributes to the Hamiltonian with
\begin{equation}
    H_{\text{island}}^n = -\mu_n \nu_z~,
\end{equation}
where $\nu_i$ are Pauli matrices acting in the electron-hole space.
For simplicity, we take the flat band limit in which each superconducting lead is described by:
\begin{equation}
    H_{SC}^i = \Delta_i \nu_x~.
\end{equation}
Finally, the hopping between the leads and the island includes effects of Rashba spin-orbit coupling and the orbital field via a Peierls phase:
\begin{equation}
    H_T^{i,n} = t_i\nu_z U_{\text{Peierls}}(\varphi_i) U_{SOC}^{i}(\alpha_{i, n})~,
\end{equation}
with
\begin{equation}
    U_{\text{Peierls}}(\varphi_i) = \exp\left(i \frac{\varphi_i}{2}\nu_z\right)~,\quad U_{SOC}^{i}(\alpha_{i,n}) = \exp\left(i\frac{\alpha_{i,n}}{2} \vec{d}_i\cdot \vec{\sigma}\right)~,
\end{equation}
where $\alpha_i$ is the spin-orbit precession angle, $\vec{d}_i$ is the direction of the spin-orbit field, $\varphi$ is the Peierls phase, and $\sigma_i$ are Pauli matrices acting in the spin space.

We arrive at the effective model by computing the self-energy corrections in the lowest orbital of the island up to fourth order in $t_i / \Delta_i$ and $t_i/\mu_1$.
We have derived the effective Hamiltonian using both the Pymablock package~\cite{rigotti_manesco_2026_18244491, day_pymablock_2024} and the analytical calculation below.
The first set of corrections corresponds to effective superconducting pairing terms in the normal island, with both second and fourth-order contributions:
\begin{equation}
    \Delta_2^{i}(\omega) = H_T^{i,0} \frac{1}{(\omega - H_{SC}^i)} (H_T^{i,0})^{\dagger}~,
\end{equation}
and
\begin{equation}
    \Delta_{4}^{ij}(\omega) = H_T^{i,0} \frac{1}{(\omega - H_{SC}^i)} \Delta_2^j(\omega) \frac{1}{(\omega- H_{SC}^i)} (H_T^{i,0})^{\dagger}~.
\end{equation}
For simplicity, we consider a resonant level, $\mu_0=0$, leading to
\begin{equation}
    \Delta_2^i(0) = -\frac{t_i^2}{\Delta_i} \nu_{\varphi_i}~,\quad \Delta_4^i(0) = \frac{t_i^2 t_j^2}{\Delta_i \Delta_j^2} \nu_{\varphi_i}~,
\end{equation}
with
\begin{equation}
    \nu_{\varphi} = \nu_x \cos \varphi + \nu_y \sin \varphi~.
\end{equation}

The second set of corrections come from closed loops involving the higher orbital:
\begin{align}
    \Sigma_4^{ij}(\omega) &= H_T^{i,0} \frac{1}{(\omega - H_{SC}^i)} (H_T^{i,1})^{\dagger} \frac{1}{(\omega - H_{\text{island}}^1)} (H_T^{j,1})^{\dagger} \frac{1}{(\omega - H_{SC}^j)} H_T^{j,0}~\\
    &= \gamma_{ij} \nu_{\varphi_i} U_{SOC}^i(\alpha_{i,0} - \alpha_{i,1})\nu_zU_{SOC}^j(\alpha_{j,1} - \alpha_{j,0})\nu_{\varphi_j}~.
\end{align}
with $\gamma_{ij} =t_i^2t_j^2/(\Delta_i\Delta_j\mu_1)$.
We fix the gauge for the spin-orbit field such that $\alpha_{i,0} = \alpha_i$, $\alpha_{i,1}=- \alpha_i$ and define $\varphi_{ij} = \varphi_i - \varphi_j$.
Thus,
\begin{align}
    \Sigma_4^{ij} &= \gamma_{ij} \nu_z\exp\left(i \varphi_{ij}\nu_z\right)\exp\left(i\alpha_{i} \vec{d}_{i}\cdot \vec{\sigma}\right)\exp\left(-i\alpha_{j} \vec{d}_{j}\cdot \vec{\sigma}\right)\\ &= \gamma_{ij} \nu_z\exp\left(i \varphi_{ij}\nu_z\right)\exp\left(i\theta_{ij} \vec{n}_{ij}\cdot \vec{\sigma}\right)~.
\end{align}
where
\begin{align}
\cos \theta_{ij} &= \cos \tilde{\alpha}_i \cos \tilde{\alpha}_j - \vec{n}_i \cdot \vec{n}_j \sin \tilde{\alpha}_i \sin \tilde{\alpha}_j~,\\
\vec{n}_{ij} \sin \theta_{ij} &= \vec{n}_i \sin \tilde{\alpha}_i \cos \tilde{\alpha}_j - \vec{n}_j \sin \tilde{\alpha}_j \cos \tilde{\alpha}_i + (\vec{n}_i \times \vec{n}_j) \sin \tilde{\alpha}_i \sin \tilde{\alpha}_j~,
\end{align}
and $\vec{n}_{i} = \vec{d}_{i}/|\vec{d}_{i}|$, $\tilde{\alpha}_{i} = |\vec{d}_{i}| \alpha_{i}$.
As a result, each pair of leads contributes
\begin{equation}
    \tilde{\Sigma}_4^{ij} = \Sigma_4^{ij}+ \Sigma_4^{ji} = \gamma_{ij}\left[\nu_z\cos \varphi_{ij} \cos \theta_{ij} + \nu_0~\vec{n}_{ij} \cdot \vec{\sigma} \sin \varphi_{ij}\sin \theta_{ij} \right]~.    
\end{equation}

The effective Hamiltonian is then
\begin{equation}
    \mathcal{H}_{\mathrm{eff}} = \Psi_0^{\dagger} H_{\mathrm{eff}} \Psi_0~,\quad H_{\mathrm{eff}} = H_{\mathrm{island}}^0 + \sum_i \left[ \Delta_2^i + \sum_{j} \left( \Delta_4^{ij}+ \Sigma_4^{ij} \right) \right]~, 
\end{equation}
from which we immediately recover the two-terminal Hamiltonian for symmetric leads:
\begin{equation}
    \sum_i \left(\Delta_2^i + \sum_j \Delta_4^{ij}\right) = \Gamma \cos \frac{\varphi}{2} \nu_x    
\end{equation}
with $\varphi = \varphi_1 - \varphi_2$, $\Gamma = -t^2 / \Delta + 2t^4 / \Delta^3$, and
\begin{equation}
    \Sigma_4^{12} + \Sigma_4^{21} = \gamma \left(
    \nu_z \cos \alpha \cos \varphi + \nu_0 \vec{n} \cdot \vec{\sigma} \sin \alpha \sin \varphi \right)~.
\end{equation}

Finally, we explicitly compute the Hamiltonian for a multiterminal device,
\begin{equation}
    H_{\mathrm{eff}} = \sum_{i,j} E_0^{ij} \cos \varphi_{ij}\nu_z + \sum_i \Gamma_i \nu_{\varphi_i} + \nu_0 \sum_{i, j} E_\sigma^{ij}~\vec{n}_{ij}\cdot\vec{\sigma} \sin \varphi_{ij}~,
\end{equation}
where $E_0^{ij} = \gamma_{ij} \cos \theta_{ij}$, $E_\sigma^{ij} = \gamma_{ij} \sin \theta_{ij}$.
Therefore, the odd-parity potential is
\begin{equation}
    U = \sum_{i,j} E_0^{ij} \cos \varphi_{ij} + \sum_{i, j} E_\sigma^{ij}~\vec{n}_{ij}\cdot\vec{\sigma} \sin \varphi_{ij}~,
\end{equation}
which are the two terms presented in Eqs.~\eqref{eq:spinless-potential} and \eqref{eq:spinful-potential}.

\section{Numerics for Solving the Circuit Hamiltonians}

We will now consider a 3-terminal ASQ shunted by capacitors, as in the circuit diagram in Fig.~\ref{fig:spinless_circuit}(a).
We start by writing the Lagrangian of this circuit:
\begin{equation}
    \mathcal{L} = \frac12 \dot{\varphi}^T C \dot{\varphi} - U(\varphi)
\end{equation}
with $\varphi^T = (\varphi_1, \varphi_2, \varphi_3)$, and
\begin{equation}
    C =
    \begin{pmatrix}
        C_{12} + C_{31} & -C_{12}           & -C_{31} \\
        -C_{12}         & C_{12} + C_{23}   & -C_{23} \\
        -C_{31}         & -C_{23}           & C_{23} + C_{31}
    \end{pmatrix}.
\end{equation}
We choose node 3 to be the reference, so the capacitance matrix reduces to
\begin{equation}
    C_{\mathrm{red}} =
    \begin{pmatrix}
        C_{12} + C_{31} & -C_{12} \\
        -C_{12}         & C_{12} + C_{23}
    \end{pmatrix}.
\end{equation}
Eliminating $\varphi_3$ from the Lagrangian, defining the canonical momenta $N_i$, and promoting the dynamical variables to quantum operators, we arrive at the following Hamiltonian:
\begin{equation}
    \mathcal{H} = \frac{4e^2}{2} \sum_{ij} (\hat{N}_i - n_{gi}) \left(C_{\mathrm{red}}\right)_{ij}^{-1} (\hat{N}_j - n_{gj}) + U(\hat{\varphi}_1, \hat{\varphi}_2)~.
\end{equation}
Finally, we define
\begin{equation}
    E_{C1} = \frac{e^2}{2} \frac{C_{12} + C_{31}}{\det C_{\mathrm{red}}}~,\quad E_{C2} = \frac{e^2}{2} \frac{C_{12} + C_{23}}{\det C_{\mathrm{red}}}~,\quad E_{C12} = e^2 \frac{C_{12}}{\det C_{\mathrm{red}}}~,
\end{equation}
so that
\begin{equation}
    \mathcal{H} = 4E_{C1}(\hat{N}_1 - n_{g1})^2 + 4E_{C2}(\hat{N}_2 - n_{g2})^2 + (\hat{N}_1 - n_{g1}) 4E_{C12} (\hat{N}_2 - n_{g2}) + U(\hat{\varphi}_1, \hat{\varphi}_2)~.
\end{equation}

We solve the Schrödinger equation numerically using Kwant~\cite{groth_kwant_2014}, discretizing the phase variables $\hat{\varphi}_i$ on a grid.
When we evaluate the spinless model, we obtain the potential $U(\varphi)$ by diagonalizing the microscopic Hamiltonian of Eq.~\eqref{eq:microscopic-model} numerically and computing:
\begin{equation}
    U(\varphi) = \frac{1}{2}\sum_{\epsilon_n < 0}\epsilon_n(\varphi) + \sum_{\epsilon_n>0} \epsilon_n(\varphi)\gamma_n^{\dagger}\gamma_n~,
    \label{eq:numerics-U}
\end{equation}
where $\gamma_n^{\dagger}$ creates a quasiparticle excitation in the groundstate.
In the main text, we use Eq.~\eqref{eq:numerics-U} to compute $U(\varphi)$ of a spinless system with one quasiparticle, \emph{i.e.} $\langle \gamma_n^{\dagger}\gamma_n\rangle =1$.
For the calculations of the spinful setup, we use the equations~\eqref{eq:spinless-potential} and~\eqref{eq:spinful-potential} directly.

\begin{figure}[!h]
    \centering
    \includegraphics[width=0.5\linewidth]{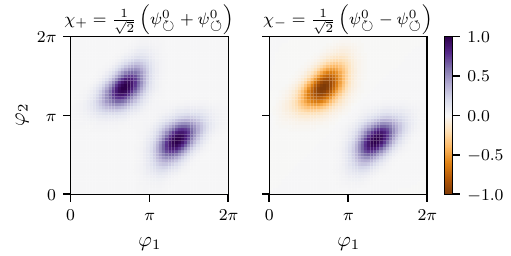}
    \caption{Two lowest eigenstates of the circuit Hamiltonian in the heavy transmon limit.
    The eigenstates are symmetric and antisymmetric linear combinations of the states located at the bottom of the two wells, $\psi_{\tau=\pm}^0$.
    We use the same parameters of Fig.~\ref{fig:spinless_circuit}}
    \label{fig:eigenstates-spinless}
\end{figure}

In the absence of spin-orbit coupling, the lowest states of the circuit Hamiltonian are located at the bottom of the potential well shown in Fig.~\ref{fig:EPRs}(c).
In each of these minima, the superconducting phases develop a chirality, which we denote by $\tau=\circlearrowright$ and $\tau=\circlearrowleft$.
The small overlap between the lowest states with opposite chirality, $\psi_{\tau}^0$, results in an exponentially small splitting between the symmetric and antisymmetric linear combinations $\chi_{\pm}$ shown in Fig.~\ref{fig:eigenstates-spinless}.

\end{document}